\documentclass[12pt]{article}

\usepackage{amsmath}
\usepackage{amssymb}
\usepackage{graphicx}
\usepackage{hyperref}
\usepackage{geometry}
\usepackage{enumitem}
\usepackage{caption}
\usepackage{subcaption}
\usepackage{cite}
\usepackage{float} 
\usepackage{tikz}
\usepackage{pgfplots}
\usepackage{subcaption}
\pgfplotsset{compat=1.18}
\usepackage{placeins}

\title{Exploring the extent of similarities in software failures across industries using LLMs}
\author{Martin Detloff \\
        \small Department of Computer Science, University of New Mexico \\
        \small \texttt{Mdetloff1@unm.edu}}
\date{\today}

\begin{document}

\maketitle

% Abstract
\begin{abstract}
\textbf{Background/Motivation}: The rapid evolution of software development necessitates enhanced safety measures. Extracting information about software failures from companies is becoming increasingly more available through news articles.

\textbf{Problem Statement}: This research utilizes the  Failure Analysis Investigation with LLMs (FAIL)
model to extract industry-specific information. Although the FAIL model's
database is rich in information, it could benefit from further categorization
and industry-specific insights to further assist software engineers. 

\textbf{Methods}: In previous work news articles were collected from reputable sources and categorized by incidents inside a database. Prompt engineering and Large Language Models (LLMs) were then applied to extract relevant information regarding the software failure. This research extends these methods by categorizing articles into specific domains and types of software failures. The results are visually represented through graphs.

\textbf{Results}: The analysis shows that throughout the database some software failures occur significantly more often in specific industries. This categorization provides a valuable resource for software engineers and companies to identify and address common failures. 

\textbf{Conclusion}: This research highlights the synergy between software engineering and Large Language Models (LLMs) to automate and enhance the analysis of software failures. By transforming data from the database into an industry specific model, we provide a valuable resource that can be used to identify common vulnerabilities, predict potential risks, and implement proactive measures for preventing software failures. Leveraging the power of the current FAIL database and data visualization, we aim to provide an avenue for safer and more secure software in the future. 
\end{abstract}

% Keywords

% Introduction
\section*{Introduction}
\label{sec:intro}
The software industry's dynamic nature demands continual advancements in safety protocols and failure analysis techniques. Current methodologies for studying software failures often face set backs due to restrictive data access and privacy concerns. The news seems to be the optimal route for collecting information using Large Language Models (LLMs). 

The FAIL database provides a well established foundation, compiling detailed reports of software failures from various news outlets. However, there exists a significant opportunity to delve deeper into the contextual specifics of these failures, such as the industries most affected and the predominant types of failures in each industry.

In response to these needs, this study expanded the use of the FAIL model by categorizing failures not just broadly but within specific industry contexts. This involved enhancing data extraction techniques through prompt engineering with LLMs, aiming to achieve a more nuanced understanding of where and why these failures occur.

Our analysis demonstrated that some types of software failures are notably more prevalent in certain industries. By identifying patterns, we provide insights that can drive targeted improvements in software practices and policies, potentially leading to substantial reductions in software failures.

% Related Work
\section*{Background And Significance}
\label{sec
}

The Securities and Exchange Commission (SEC) recently made a ruling requiring public companies to disclose cybersecurity-related failures. [7] This means that any breach in security deemed to have a significant impact or potential harm must be reported within four business days of the incident. Public companies must also create annual reports of cybersecurity risk management strategies and governance. With this increased transparency requirement in the industry, the FAIL model and industry-specific insights will only become more useful and prevalent in the future.

Previous research has laid the foundation for using Large Language Models (LLMs) to analyze software failures. In "Can Large Language Models Analyze Software Failures in the News? An End-to-End Automated Pipeline with FAIL," [1] Anandayuvaraj demonstrated the potential of LLMs to automate the process of analyzing news articles and extracting relevant information about software failures. This work shows the insurmountable importance of the use of LLMs in software failure analysis. With this study many different avenues open in terms of research such as why software failures happen, methods of prevention, and common software failure risks. 

Another significant study, "Reflecting on Recurring Failures in IoT Development," [11], highlighted the challenges specific to Internet of Things (IoT) Development. Anandayuvaraj identified common patterns of failures in IoT systems and emphasized the importance of incorporating lessons learned into future designs to prevent recurring consistent issues. This research not only enforces the necessity of industry-specific analysis to address unique challenges, but also supports assumptions that some failures could be dominant in software development.

In "Incorporating Failure Knowledge into Design Decisions for IoT Systems: A Controlled Experiment on Novices," [12] Anandayuvaraj explored how failure knowledge could be integrated into the design process. This study focused on evaluating the influence of failure-related learning techniques. The study revealed that providing novice developers with insights into common failures can significantly enhance the quality of IoT systems. This experiment supports the idea that sharing industry-specific failure knowledge can improve software design and development.

In "Reflections on Software Failure Analysis," [14] Amusuo explored different software failure studies along with common shortcomings of these previous studies. They found that studies have a bias towards open source data, and there is a lack of consistency in taxonomies used in identifying failures. This study highlights the importance of a universal taxonomy for comparison throughout different studies, as well as a need for improved replicability amongst studies. 

Finally, "An Empirical Study on Using Large Language Models to Analyze Software Supply Chain Security Failures" [13] examined the applicability of LLMs in identifying vulnerabilities in the software supply chain. Anandayuvaraj's findings highlighted the effectiveness, and some possible drawbacks of LLMs in identifying patterns and potential risks within supply chain security.

These works collectively illustrate the potential of leveraging LLMs and industry-specific insights in understanding and preventing software failures. By building upon this foundation, the current research aims to further refine these methodologies and enhance the utility of the FAIL model in delivering insights for software engineers and companies.

\section*{Related Work}
\label{sec:related}
This work is directly correlated to the work done by FAIL: Analyzing Software Failures from the News Using LLMs [1], materializing a way to automate the process of gathering information on software failures from news articles. In this paper they explored an efficient approach to automate the process of collecting information on recent software failures from the news. They collected information only from articles that the LLMs deemed to be having sufficient enough information to analyze. They then categorized this data into summaries, postmortems, and taxonomy. They collected valuable information, and the information is informative. The information could benefit from further insights into industry-specific failures. 

% Methodology
\section*{Methodology}
\label{sec:method}

We began by creating a prompt to collect data from the database and categorize it into industry-specific failures. The initial taxonomy was influenced by previous works [8, 9, 10], which provide frameworks for categorizing software failures. To increase consistency, we manually set the temperature of the LLM (chatgpt-3.5-turbo) to 0.

\subsection*{Initial Prompt Design}

The initial prompt included general instructions such as:

\begin{verbatim}
You are an AI trained to identify the type of software failure based 
on the Cause provided. Use the context in the cause to determine the
type of software failure
\end{verbatim}

However, this prompt's performance was suboptimal, as effective prompts need to be clear and concise to guide the Large Language Model (LLM) in producing the desired output [2]. The initial prompt generated results with high variability or "randomness."

\subsection*{Iteration 1: Specific Failure List}

To improve accuracy, we revised the prompt with a specific list of common software failures, manually curated based on descriptions in the database and taxonomies widely used in the industry for categorizing software failures [10, 9]. The revised prompt was:

\begin{verbatim}
You are an AI trained to identify the type of software failure based on 
the Cause provided. Use the context in the cause to determine the type
of software failure from the following list:

        - Data Breach 
        - Functionality Bug 
        - UI/UX Bug  
        - Regression Bug 
        - Outage 
        - Security Vulnerability 
        - Performance Issue 
        - Integration Issue 
        - Non-Software Cause  
        - Other

Your task is to read the cause below and determine the type of software 
failure from the given list. Only return the type of failure without any 
additional text or explanation. If the cause does not fit any category 
in the list, return "Other".
\end{verbatim}

This adjustment yielded more consistent results, but some incorrect categorizations persisted.

\subsection*{Iteration 2: Few-Shot Prompting}

To further refine the prompt, we incorporated few-shot prompting, providing examples of correct categorizations. The updated prompt was:

\begin{verbatim}
You are an AI trained to identify the type of software failure based on 
the Cause provided. Use the context in the cause to determine the type
of software failure from the following list:

        - Data Breach 
        - Functionality Bug 
        - UI/UX Bug  
        - Regression Bug 
        - Outage 
        - Security Vulnerability 
        - Performance Issue 
        - Integration Issue 
        - Non-Software Cause  
        - Other
        
Your task is to read the cause below and determine the type of software 
failure from the given list. Only return the type of failure without any 
additional text or explanation. If the cause does not fit any category 
in the list, return "Other"

Examples:

Cause: The faulty computer software in the Chinook helicopter was described
as "positively dangerous" and had deficiencies that meant the pilot's full 
control of the engines "could not be assured". 
Functionality Bug

Cause: Software bug causing Windows Mobile devices to incorrectly date
incoming SMS messages to 2016.
Regression Bug

...

Cause: Lack of Adobe Flash support on Apple's Safari browser for iPad, iPhone,
and iPod Touch devices.
Non-Software Cause

\end{verbatim}

This iteration significantly improved the prompt's performance.

\subsection*{Evaluation and Results}

\begin{figure}[H]
    \centering
    \includegraphics[width=0.75\linewidth]{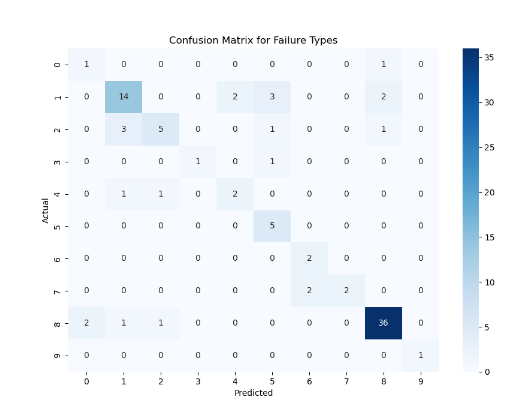}
    \caption{Confusion matrix used for analysis.}
    \label{fig:enter-label}
\end{figure}

We evaluated the prompt's accuracy using a confusion matrix and a small dataset of 90 articles. The overall average accuracy of the prompt was 76 percent (reference Figure \ref{fig:enter-label} for an example of the matrix used). With this accuracy, we ran the script through the database to categorize each software failure by industry. Results are shown visually through bar graphs.
Explanation
% Results
\section*{Results}
\label{sec:results}

\subsection*{Overview}
This section presents the analysis of software failure frequencies across different industries.

\subsubsection*{Finance Industry}

The graph in figure \ref{fig:finance_industry} shows that security-related failures are the most common in the finance industry, with Security vulnerabilities being significantly more frequent than other types of failures.

\subsubsection*{Healthcare Industry}
As shown in figure \ref{fig:healthcare_industry} shows security vulnerabilities are the most common in the healthcare industry. This could be due to the healthcare industry storing a vast amount of patient data with inadequate security. 

\subsubsection*{Information Industry}
As shown in figure \ref{fig:Information_industry} shows security vulnerabilities are most common in the information industry. The vast amount of data within the IT (information Technology) industry increases the risk of security related failures.  

\subsubsection*{Knowledge Industry}
As shown in figure \ref{fig:knowledge_industry} shows security vulnerabilities are most common in the Knowledge industry. This industry includes educational institutions, research organizations, and knowledge-based enterprises. Security related failures could be happening more frequently due to the openness of sharing information in an educational environment.

\subsubsection*{Transportation Industry}
As shown in figure \ref{fig:transportation_industry} shows functionality bugs are most common in the Transportation industry. This industry relies heavily on complex logistics and random variables, which could in effect be the reason we are seeing more functionality bugs.

\subsubsection*{Entertainment Industry}
\begin{figure}[H]
\centering
\caption{Industry Specific Results}

\begin{subfigure}{.45\textwidth}
\centering
\begin{tikzpicture}
\begin{axis}[
    width=\linewidth,
    height=4.5cm,
    ybar,
    enlargelimits=0.15,
    ylabel={Number of failures},
    symbolic x coords={Security Vulnerability,Functionality Bug,Data Breach,Outage,Integration Issue,Other,Performance Issue,UI/UX Bug,Regression Bug,Non-Software Cause},
    xtick=data,
    nodes near coords,
    every node near coord/.append style={font=\tiny},
    x tick label style={rotate=45,anchor=east},
    bar width=7pt,
    xlabel={Failure Type},
    ylabel style={align=center},
    tick label style={font=\tiny},
    title={In Finance Industry},
]
\addplot[fill=blue!60] coordinates {
    (Security Vulnerability,95)
    (Functionality Bug,11)
    (Data Breach,35)
    (Outage,24)
    (Integration Issue,12)
    (Other,5)
    (Performance Issue,4)
    (UI/UX Bug, 0)
    (Regression Bug, 0)
    (Non-Software Cause,3)
};
\end{axis}
\end{tikzpicture}
\caption{Graph of failures within the finance industry}
\label{fig:finance_industry}
\end{subfigure}%
\hfill
\begin{subfigure}{.45\textwidth}
\centering
\begin{tikzpicture}
\begin{axis}[
    width=\linewidth,
    height=4.5cm,
    ybar,
    enlargelimits=0.15,
    ylabel={Number of failures},
    symbolic x coords={Security Vulnerability,Functionality Bug,Data Breach,Outage,Integration Issue,Other,Performance Issue,UI/UX Bug,Regression Bug,Non-Software Cause},
    xtick=data,
    nodes near coords,
    every node near coord/.append style={font=\tiny},
    x tick label style={rotate=45,anchor=east},
    bar width=7pt,
    xlabel={Failure Type},
    ylabel style={align=center},
    tick label style={font=\tiny},
    title={In Health Industry},
]
\addplot[fill=blue!60] coordinates {
    (Security Vulnerability,47)
    (Functionality Bug,20)
    (Data Breach,19)
    (Outage,6)
    (Integration Issue,6)
    (Other,5)
    (Performance Issue,5)
    (UI/UX Bug,2)
    (Regression Bug, 0)
    (Non-Software Cause,1)
};
\end{axis}
\end{tikzpicture}
\caption{Graph of failures within the healthcare industry}
\label{fig:healthcare_industry}
\end{subfigure}%

\vspace{1em}

\begin{subfigure}{.45\textwidth}
\centering
\begin{tikzpicture}
\begin{axis}[
    width=\linewidth,
    height=4.5cm,
    ybar,
    enlargelimits=0.15,
    ylabel={Number of failures},
    symbolic x coords={Security Vulnerability,Functionality Bug,Data Breach,Outage,Integration Issue, Other,Performance Issue,UI/UX Bug,Regression Bug, Non-Software Cause},
    xtick=data,
    nodes near coords,
    every node near coord/.append style={font=\tiny},
    x tick label style={rotate=45,anchor=east},
    bar width=7pt,
    xlabel={Failure Type},
    ylabel style={align=center},
    tick label style={font=\tiny},
    title={In Information Industry},
]
\addplot[fill=blue!60] coordinates {
    (Security Vulnerability,716)
    (Functionality Bug,127)
    (Data Breach,120)
    (Outage,64)
    (Integration Issue,33)
    (Other,31)
    (Performance Issue,26)
    (UI/UX Bug,18)
    (Regression Bug,16)
    (Non-Software Cause, 5)
};
\end{axis}
\end{tikzpicture}
\caption{Graph of failures within the information industry}
\label{fig:Information_industry}
\end{subfigure}%
\hfill
\begin{subfigure}{.45\textwidth}
\centering
\begin{tikzpicture}
\begin{axis}[
    width=\linewidth,
    height=4.5cm,
    ybar,
    enlargelimits=0.15,
    ylabel={Number of failures},
    symbolic x coords={Security Vulnerability,Functionality Bug,Data Breach,Outage,Integration Issue,Other,Performance Issue,UI/UX Bug,Regression Bug,Non-Software Cause},
    xtick=data,
    nodes near coords,
    every node near coord/.append style={font=\tiny},
    x tick label style={rotate=45,anchor=east},
    bar width=7pt,
    xlabel={Failure Type},
    ylabel style={align=center},
    tick label style={font=\tiny},
    title={In Knowledge Industry},
]
\addplot[fill=blue!60] coordinates {
    (Security Vulnerability,18)
    (Functionality Bug,12)
    (Data Breach,5)
    (Outage,3)
    (Integration Issue,8)
    (Other,14)
    (Performance Issue,6)
    (UI/UX Bug, 0)
    (Regression Bug, 0)
    (Non-Software Cause, 0)
};
\end{axis}
\end{tikzpicture}
\caption{Graph of failures within the knowledge industry}
\label{fig:knowledge_industry}
\end{subfigure}%

\vspace{1em}

\begin{subfigure}{.45\textwidth}
\centering
\begin{tikzpicture}
\begin{axis}[
    width=\linewidth,
    height=4.5cm,
    ybar,
    enlargelimits=0.15,
    ylabel={Number of failures},
    symbolic x coords={ Security Vulnerability,Functionality Bug,Data Breach,Outage,Integration Issue,Other,Performance Issue,UI/UX Bug,Regression Bug,Non-Software Cause },
    xtick=data,
    nodes near coords,
    every node near coord/.append style={font=\tiny},
    x tick label style={rotate=45,anchor=east},
    bar width=7pt,
    xlabel={Failure Type},
    ylabel style={align=center},
    tick label style={font=\tiny},
    title={In Transportation Industry},
]
\addplot[fill=blue!60] coordinates {
    (Security Vulnerability,80)
    (Functionality Bug,187)
    (Data Breach,4)
    (Outage,16)
    (Integration Issue,57)
    (Other,52)
    (Performance Issue,9)
    (UI/UX Bug,6)
    (Regression Bug,4)
    (Non-Software Cause,2)
};
\end{axis}
\end{tikzpicture}
\caption{Graph of failures within the transportation industry}
\label{fig:transportation_industry}
\end{subfigure}%
\hfill
\begin{subfigure}{.45\textwidth}
\centering
\begin{tikzpicture}
\begin{axis}[
    width=\linewidth,
    height=4.5cm,
    ybar,
    enlargelimits=0.15,
    ylabel={Number of failures},
    symbolic x coords={Security Vulnerability,Functionality Bug,Data Breach,Outage,Integration Issue,Other,Performance Issue,UI/UX Bug,Regression Bug,Non-Software Cause},
    xtick=data,
    nodes near coords,
    every node near coord/.append style={font=\tiny},
    x tick label style={rotate=45,anchor=east},
    bar width=7pt,
    xlabel={Failure Type},
    ylabel style={align=center},
    tick label style={font=\tiny},
    title={In Entertainment Industry},
]
\addplot[fill=blue!60] coordinates {
    (Security Vulnerability,60)
    (Functionality Bug, 33)
    (Data Breach,24)
    (Outage,4)
    (Integration Issue,9)
    (Other,6)
    (Performance Issue,15)
    (UI/UX Bug,4)
    (Regression Bug, 0)
    (Non-Software Cause, 0)
};
\end{axis}
\end{tikzpicture}
\caption{Graph of failures within the entertainment industry}
\label{fig:entertainment_industry}
\end{subfigure}%

\vspace{1em}
\clearpage

\end{figure}
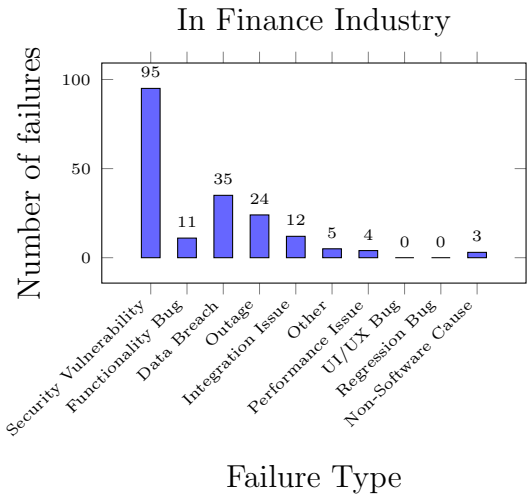

As shown in figure \ref{fig:entertainment_industry} shows security vulnerabilities are most common in the Entertainment industry. This industry has a large audience, and often uses third party software to manage data, which could lead to increased vulnerabilities.

\subsubsection*{Government Sector}
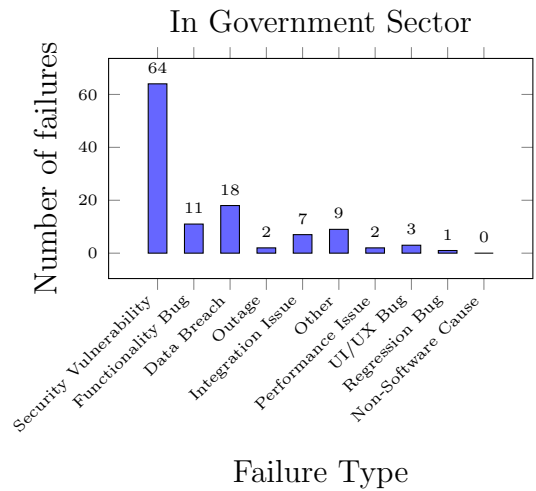
\begin{figure}[!htbp]
\centering
\caption{Industry Specific Results}
\begin{subfigure}{.45\textwidth}
    
\begin{tikzpicture}
\begin{axis}[
    width=\linewidth,
    height=4.5cm,
    ybar,
    enlargelimits=0.15,
    ylabel={Number of failures},
    symbolic x coords={Security Vulnerability,Functionality Bug,Data Breach,Outage,Integration Issue,Other,Performance Issue,UI/UX Bug,Regression Bug,Non-Software Cause},
    xtick=data,
    nodes near coords,
    every node near coord/.append style={font=\tiny},
    x tick label style={rotate=45,anchor=east},
    bar width=7pt,
    xlabel={Failure Type},
    ylabel style={align=center},
    tick label style={font=\tiny},
    title={In Government Sector},
]
\addplot[fill=blue!60] coordinates {
    (Security Vulnerability,64)
    (Functionality Bug, 11)
    (Data Breach,18)
    (Outage,2)
    (Integration Issue,7)
    (Other,9)
    (Performance Issue,2)
    (UI/UX Bug,3)
    (Regression Bug, 1)
    (Non-Software Cause,0)
};
\end{axis}
\end{tikzpicture}
\caption{Graph of failures within the government sector}
\label{fig:government_industry}
\end{subfigure}
\end{figure}

As shown in figure \ref{fig:government_industry} shows security vulnerabilities are most common in the Government Sector. With the government sector handling a large range of sensitive information, it's no surprise that security vulnerabilities are at the forefront of software failures for this domain. 
% \begin{figure} % H option enforces the placement of the figure
%     \centering
%     \includegraphics[width=0.75\textwidth]{999.png}
%     \caption{HeatMap of software failures in every industry.}
%     \label{fig:every_industry}
% \end{figure}
% \subsection*{Frequency of Software Failures by Industry}

% Discussion
\section*{Discussion}
\label{sec:discussion}
\subsection*{Interpretation of Results:}
Our results indicate that certain software failures occur significantly more frequently in specific industries. For instance, the analysis revealed that the information industry experiences a higher incidence of security-related failures, while the transportation industry often encounters functionality issues. These findings highlight the distinct challenges faced by different sectors, underscoring the need for tailored approaches to software reliability. We can also see a trend across industries, suggesting that many different industries are facing the same software development challenges.

Our study dives deeper by quantifying these differences and providing a detailed breakdown of failure types within each industry. This granularity offers new insights into the specific vulnerabilities and common issues faced by different sectors.

\subsection*{Implications of Findings }
\begin{itemize}
    \item \textbf{Industry specific insights:} Our study provides industry-specific insights into software failures to better cater efforts towards a specific industry. 
    \item \textbf{Efficiency: } This could allow for more efficient software maintenance and improvements when a lack of resources is present. 
\end{itemize}
% Policy and Best Practices
\subsection*{Policy and Best Practices}
\label{sec:policy}

Given that security vulnerabilities are the most frequent software failure across the large majority of industries, we recommend the following policy and best practices based upon studies such as [3, 4, 5]:

\begin{enumerate}
    \item \textbf{Regular Security Audits:}
    Conduct regular security audits to identify and mitigate vulnerabilities in the software. This includes code reviews, penetration testing, and compliance checks. Regular security audits can aid in catching security vulnerabilities early when a fix is more plausible. 
    
    \item \textbf{Security Training:}
    Provide ongoing security training for software engineers and developers to stay updated with the latest security threats and mitigation techniques. Informing the group of developers about common security breaches and fixes in the past can be helpful in preventing them in the future.  
    
    \item \textbf{Secure Coding Practices:}
    Implement secure coding practices, such as input validation, error handling, and secure authentication mechanisms, to prevent common vulnerabilities like SQL injection and cross-site scripting (XSS). Making sure that your coding practices are consistent, and secure is one of the most practical solutions to security vulnerabilities, as we realize software development can involve large teams all working in separate sectors.  Consistency can benefit debugging, and communication pertaining to security.
    
    \item \textbf{Incident Response Plan:}
    Develop and maintain a comprehensive incident response plan to quickly address and mitigate the impact of security breaches. Having a response plan will mitigate damages, and ensure that the security vulnerabilities are dealt with quickly and efficiently. 
    
    \item \textbf{Access Control:}
    Ensure strict access control policies are in place, with the principle of least privilege, to minimize the risk of unauthorized access to sensitive data. For more information please see [4].
    
    \item \textbf{Regular Updates:}
    Keep all software and dependencies updated with the latest security patches and updates to protect against known vulnerabilities.
    
\end{enumerate}
\subsubsection*{Limitations: }
\begin{itemize}
    \item \textbf{Limited scope: } Our study uses only the data found in the FAIL database, which does not capture every failure across every industry and can be seen as limited.
    \item \textbf{Use of Large Language Models: } Our study utilizes Large Language Models to categorize software failures. Large Language models can be inaccurate, and have the possibility of giving false information known as hallucinations.  
    \item  \textbf{Using News as a Source: } The FAIL database uses news articles as a source of collecting information on software failures. News can have biases, and have only bits of information. 
\end{itemize}

% Conclusion
\section*{Conclusion}
\label{sec:conclusion}
In conclusion our study highlights the importance of industry-specific software failure distinctions. By identifying the most common failures in each specific sector we can inform more directive preventive measures towards software failures. These findings have significant implications for software engineers and industry best practices, paving the way for a safer software and practices in the future.

Our research provides a foundation for future research to build upon, with the potential to enhance the FAIL model with new tools, and better informational gathering techniques. Ultimately our work contributes to a deeper understanding of software failures and supports the development of targeted solutions to prevent industry-specific software failures in the future. 

% Acknowledgements
\section*{Acknowledgements}
\begin{itemize}
    \item PI: Jamie C Davis: For his invaluable guidance, support, and encouragement throughout this research project. 
    \item This project was funded by NSF award CNS-1836952 via the SIRI program
    \item Peer: Bryan Zhang for useful feedback/reviews during the creation of this study
\end{itemize}
% References
\bibliographystyle{plain}

\section*{References}
[1] Anandayuvaraj, Dharun, et al. "Can Large Language Models Analyze Software Failures in the News? An End-to-End Automated Pipeline with FAIL." arXiv preprint arXiv:2406.08221 (2024).
\newline \newline
[2] Giray, L. Prompt Engineering with ChatGPT: A Guide for Academic Writers. Ann Biomed Eng 51, 2629–2633 (2023). https://doi.org/10.1007/s10439-023-03272-4 
\newline \newline
[3] R. N. Charette, "Why software fails ," in IEEE Spectrum, vol. 42, no. 9, pp. 42-49, Sept. 2005, doi: 10.1109/MSPEC.2005.1502528.
keywords: {Business;Investments;Merchandise;
Computer crashes;FAA;Information technology;Programmable control;History;Companies;Costs},
\newline \newline
[4] Lutkevich, Ben. “What Is Access Control?” Security, TechTarget, 7 July 2022, www.techtarget.com/searchsecurity/definition/access-control. 
\newline \newline
[5] Greenberg, Andy. “The Full Story of the Stunning RSA Hack Can Finally Be Told.” Wired, Conde Nast, 20 May 2021, www.wired.com/story/the-full-story-of-the-stunning-rsa-hack-can-finally-be-told/. 
\newline \newline
[6] Etheredge, Justin. “What’s up with All of These Software Security Vulnerabilities?” Simple Thread, 29 Dec. 2021, www.simplethread.com/whats-up-with-all-of-these-software-security-vulnerabilities/. 
\newline \newline
[7] Securities, U. S., \& Exchange Commission. (2023). SEC adopts rules on cybersecurity risk management, strategy, governance, and incident disclosure by public companies. press release, (2023-139), 2023-139.
\newline \newline
[8] Engineering Standards Committee of the IEEE Computer Society (2009) IEEE Std 1044-2009 (Revision of IEEE Std 1044-1993), IEEE Standard Classification for Software Anomalies.
\newline \newline
[9] Hamill M, Goseva-Popstojanova K (2015) Exploring fault types, detection activities, and failure severity in an evolving safety-critical software system. Software Qual J 23:229–265.
\newline \newline
[10] Mariani L (2003) A fault taxonomy for component-based software. Electron Notes Theoret Comput Sci 82(6):55–65.
\newline \newline
[11] Anandayuvaraj, D., Hossain, M., \& Saha, D. (2023). "Reflecting on Recurring Failures in IoT Development" Proceedings of the 36th IEEE/ACM International Conference on Automated Software Engineering (ASE '23), 120-131. https://doi.org/10.1145/3551349.
3559545
\newline \newline
[12] D. Anandayuvaraj, P. Thulluri, J. Figueroa, H. Shandilya and J. C. Davis, "Incorporating Failure Knowledge into Design Decisions for IoT Systems: A Controlled Experiment on Novices," 2023 IEEE/ACM 5th International Workshop on Software Engineering Research and Practices for the IoT (SERP4IoT), Melbourne, Australia, 2023, pp. 33-37, doi: 10.1109/SERP4IoT59158.2023.00011. keywords: {Software design;Design methodology;Control systems;Software;Safety;Internet of Things;Engineering students;Software Engineering;Internet of Things;IoT}
\newline \newline
[13] Tanmay Singla, Dharun Anandayuvaraj, Kelechi G. Kalu, Taylor R. Schorlemmer, and James C. Davis. 2023. An Empirical Study on Using Large Language Models to Analyze Software Supply Chain Security Failures. In Proceedings of the 2023 Workshop on Software Supply Chain Offensive Research and Ecosystem Defenses (SCORED '23). Association for Computing Machinery, New York, NY, USA, 5–15. https://doi.org/10.1145/
3605770.3625214
\newline \newline
[14] Paschal C. Amusuo, Aishwarya Sharma, Siddharth R. Rao, Abbey Vincent, and James C. Davis. 2022. Reflections on software failure analysis. In Proceedings of the 30th ACM Joint European Software Engineering Conference and Symposium on the Foundations of Software Engineering (ESEC/FSE 2022). Association for Computing Machinery, New York, NY, USA, 1615–1620. https://doi.org/10.1145/3540250.3560879

\end{document}